# Comment on "Broken translational and rotational symmetry via charge stripe order in underdoped YBa$_2$Cu$_3$O$_{6+y}$"

Author: B. V. Fine[1,2,*]

**Affiliations:**

[1] Skolkovo Institute of Science and Technology, 100 Novaya Str., Skolkovo, Moscow Region 143025, Russia

[2] Institute for Theoretical Physics, University of Heidelberg, Philosophenweg 12, 69120 Heidelberg, Germany

* Email: b.fine@skoltech.ru

**Abstract**: Comin *et al.* [Science **347**, 1335 (2015)] have interpreted their resonant X-ray scattering experiment as indicating that charge inhomogeneities in the family of high-temperature superconductors YBa$_2$Cu$_3$O$_{6+y}$ (YBCO) have the character of one-dimensional stripes rather than two-dimensional checkerboards. The present comment shows that one cannot distinguish between stripes and checkerboards on the basis of the above experiment.

Comin *et al.* [1] conducted resonant X-ray scattering (RXS) experiment for three different compounds belonging to the family of high-temperature superconductors YBa$_2$Cu$_3$O$_{6+y}$ (YBCO). The experiment focused on the accurate measurements of the shapes of four charge ordering peaks appearing in the RXS structure factor $S(q_x, q_y)$ at positions $\{\pm Q_0, 0\}$ and $\{0, \pm Q_0\}$, where $Q_0 \approx \pi/2$ in the units of inverse lattice period. These peaks can originate from either two-dimensional (2D) checkerboard-like modulations of charge density or 1D stripe-like modulations. The stripe interpretation implies that the sample can be fully partitioned into regions of mutually orthogonal 1D modulations. Each such a region would generate only two dominant peaks at either $\{\pm Q_0, 0\}$ or $\{0, \pm Q_0\}$. The checkerboard interpretation implies that, under any partition, there will be regions generating all four peaks with comparable intensity. Distinguishing between stripes and checkerboards was the goal of Comin *et al.* In other families of high-temperature superconductors, this goal proved to be notoriously difficult to achieve, see e.g. [2,3,4,5]. In particular, the experimental effort of Ref.[3] and the discussion given in Ref.[4] are very reminiscent of the present case.

Comin *et al.* have found that the shapes of measured charge peaks are elongated in the direction perpendicular to the wave vectors defining the centers of the peaks. In their analysis, the above authors associated the finite width of the peaks with the finite size of either stripe or checkerboard domains and then, in the main text of the article, they pointed out that, in the case of checkerboards, the shapes of individual peaks should either have the four-fold symmetry, i.e. be not elongated, or the orientation of the elongation should be different. At the same time, stripe domains could reproduce the observed elongated peak shapes. The supplementary material, however, indicated that the difference was not so clear-cut, because "canted" checkerboard domains would reproduce the observed elongation of the charge peaks as well. Comin *et al.* then introduced a quantitative constraint on the canted checkerboard scenario (equation S16 of their Supplementary Material), and, in supplementary table S3, showed that their experimental results violate this constraint. Finally, Comin *et al.* concluded that their experimental observations are incompatible with checkerboard modulations and hence indicate stripe-like modulations.

The goal of this comment is to raise the objection to the above conclusion. The problem with the reasoning of Comin *et al.* is that instead of doing the Fourier analysis directly, they adopted various oversimplifying assumptions involving rigid domains of perfectly periodic structures for both stripes and checkerboards. Adopting a domain picture amounts to an implicit assumption of a certain kind of phase coherence between different Fourier components of the charge modulation, for which, to the best of this author's knowledge, there is no experimental evidence. At the same time, in the opposite limit of no coherence between different modulation harmonics, sufficiently narrow charge peaks, such as those observed by Comin *et al.*, are consistent with a rather routinely looking checkerboard modulation *irrespective of the shape of the peaks*.

To demonstrate the above statement, let us assume Gaussian peak shapes (which will not be essential for the conclusions) and represent the four peaks in the structure factor as

$$S(q_x, q_y) = \sum_{i=1}^{4} A_i \exp\left\{ -\frac{(q_x - Q_{x,i})^2}{2\sigma_{x,i}^2} - \frac{(q_y - Q_{y,i})^2}{2\sigma_{y,i}^2} \right\},$$

where $i$ is the index that of the peaks, $\{Q_{x,i}, Q_{y,i}\}$ the positions of the centers of the peaks admitting values $\{Q_0, 0\}$, $\{-Q_0, 0\}$, $\{0, Q_0\}$, $\{0, -Q_0\}$; $\sigma_{x,i}, \sigma_{y,i}$ are the peak widths in the directions indicated by the subscripts, and $A_i = \frac{1}{8\pi\sigma_{x,i}\sigma_{y,i}}$ are the normalization constants.

Let us choose $Q_0 = \pi/2$ and then $\sigma_{x,i} = 0.085 Q_0$, $\sigma_{y,i} = 0.2 Q_0$ for peaks centered at $\{Q_0, 0\}$ and $\{-Q_0, 0\}$, and $\sigma_{x,i} = 0.15 Q_0$, $\sigma_{y,i} = 0.1 Q_0$ for peaks centered at $\{0, Q_0\}$ and $\{0, -Q_0\}$. The resulting structure factor $S(q_x, q_y)$ is plotted in Fig.1A. Such a choice of parameters violates the constraint (equation S16 of Ref.[1]) on the canted checkerboards by factor of about 2. In the framework of the assumptions adopted in Ref.[1], these four peaks cannot correspond to a checkerboard.

Figure 1B demonstrates the two-dimensional Fourier transforms of the above structure factor, which gives the correlation function $C(x, y) \cong \langle \delta\rho(x + x_0, y + y_0) \delta\rho(x_0, y_0) \rangle_{x_0, y_0}$, where $\delta\rho(x, y)$ describes the fluctuation of the charge density with respect to the average value, and $\langle ... \rangle_{x_0, y_0}$ denotes averaging over $x_0$ and $y_0$. Independently of the shape of the four narrow peaks, $C(x, y)$ is bound to show strong checkerboard correlations of the kind appearing in Fig.1B.

Now, to generate a possible pattern of two-dimensional density fluctuations $\delta\rho(x, y)$ underlying the peaks in $S(q_x, q_y)$, let us recall that the $S(q_x, q_y)$ also represents the square of the amplitude of a harmonic with wave numbers $\{q_x, q_y\}$. Therefore, for the sake of producing an example, let us resolve the spectral peaks around $\{Q_0, 0\}$ and $\{0, Q_0\}$ into two 41×41 grids of discrete Fourier components uniformly spanning the ranges $[-4\sigma_{x,i}, 4\sigma_{x,i}]$ and $[-4\sigma_{y,i}, 4\sigma_{y,i}]$ around the peak centers in the *x*- and *y*-directions respectively. [The apparent character of the resulting function $\delta\rho(x, y)$ does not change, if more dense $\{q_x, q_y\}$ grid is used.] The density modulations in the real space are then obtained as

$$\delta\rho(x, y) = \sum_m a_m \cos(q_{x,m} x + q_{y,m} y + \varphi_m),$$

where index *m* labels all 2×41×41 discrete Fourier components participating in the expansion, $a_m = \sqrt{S(q_{x,m}, q_{y,m})}$ the corresponding amplitude (up to a normalization constant), and $\varphi_m$ is the random phase of each component.

The numerically generated function $\delta\rho(x,y)$ for one possible set of random phases $\varphi_m$ is shown in Fig.1C. It conveys a clear impression of a fluctuating checkerboard, in fact not much different from the results of the scanning tunneling microscopy experiments [6,7] for other cuprate compounds. As seen in Fig.1C, the randomness of the phases $\varphi_m$ implies that the correlation length controlling the width of the charge modulation peaks originates from the distortions of mostly continuous superstructures rather than from the domains of perfectly periodic modulations.

To conclude, the experimental results reported in Ref.[1] represent new valuable microscopic information that has implications for both the stripe and the checkerboard scenarios of charge modulations in cuprates. As such, however, the above results do not rule out the checkerboard modulations.

**Acknowledgements:**

I am grateful to V. K. Bhartiya for drawing my attention to the subject of this comment.

**Figures:**

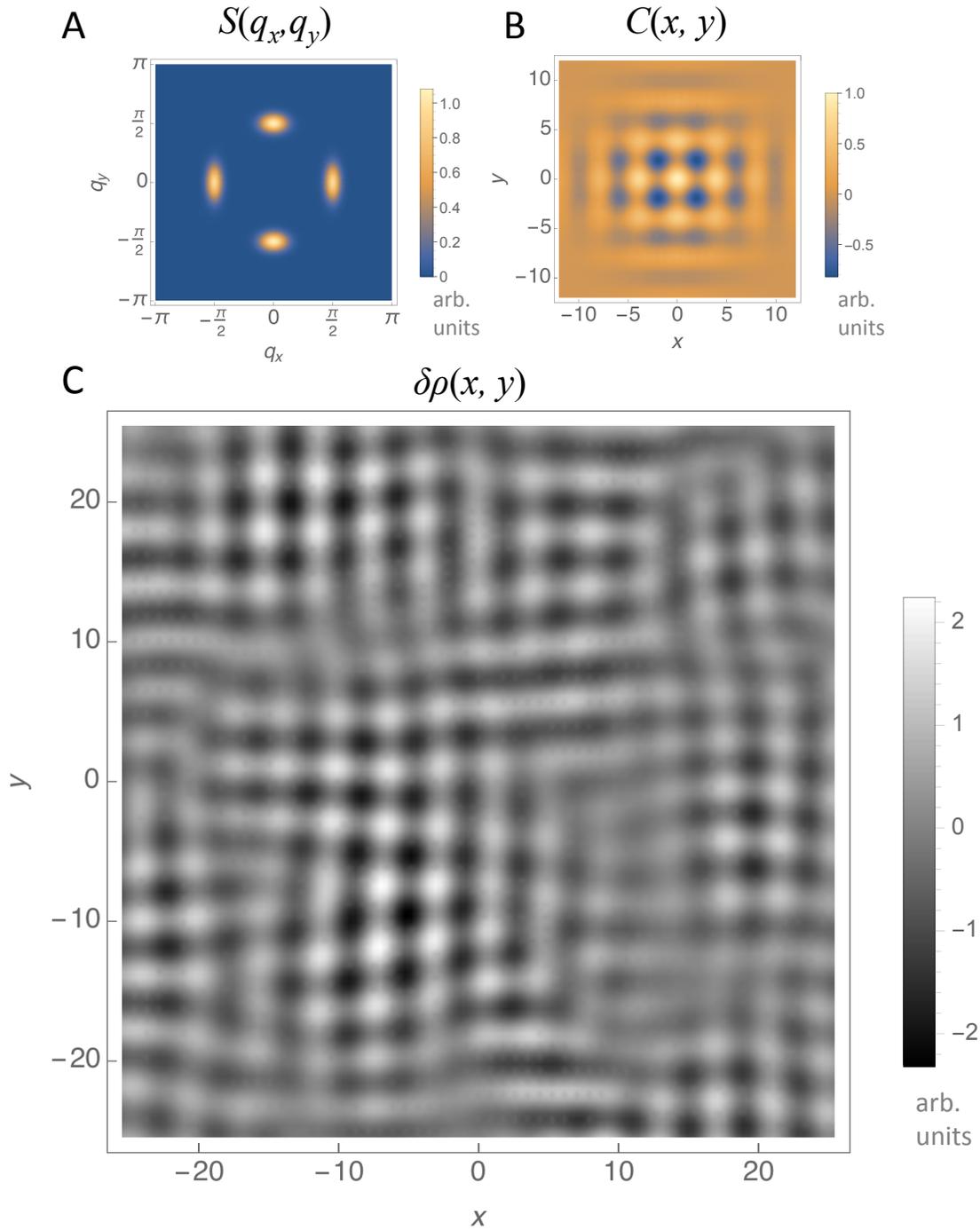

**Fig.1. Fluctuating checkerboard modulation of charge density.** (**A**) Four charge modulation peaks in the structure factor $S(q_x, q_y)$ with parameters given in the text. (**B**) Density-density correlation function $C(x, y)$ obtained as the two-dimensional Fourier transform of $S(q_x, q_y)$ given in (A). (**C**) Example of density modulation $\delta\rho(x, y)$ corresponding to $S(q_x, q_y)$ given in (A).